\documentclass[12pt,preprint]{aastex}

\shorttitle{Tsunamis in Galaxy Clusters}
\shortauthors{Fujita, Suzuki, \& Wada}

\begin{document}

\title{Tsunamis in Galaxy Clusters: Heating of Cool Cores
by Acoustic Waves}

\author{Yutaka Fujita\altaffilmark{1,2}, Takeru Ken
Suzuki\altaffilmark{3,4}, and Keiichi Wada\altaffilmark{1}}

\altaffiltext{1}{National Astronomical
Observatory, Osawa 2-21-1, Mitaka, Tokyo 181-8588, Japan}
\email{yfujita@th.nao.ac.jp}
\altaffiltext{2}{Department
of Astronomical Science, The Graduate University for Advanced Studies,
Osawa 2-21-1, Mitaka, Tokyo 181-8588, Japan}
\altaffiltext{3}{Department of Physics, Kyoto University, Kitashirakawa,
Sakyo-ku, Kyoto 606-8502, Japan}
\altaffiltext{4}{JSPS Research Fellow}

\begin{abstract}
 Using an analytical model and numerical simulations, we show that
 acoustic waves generated by turbulent motion in intracluster medium
 effectively heat the central region of a so-called ``cooling flow''
 cluster. We assume that the turbulence is generated by substructure
 motion in a cluster or cluster mergers. Our analytical model can
 reproduce observed density and temperature profiles of a few
 clusters. We also show that waves can transfer more energy from the
 outer region of a cluster than thermal conduction alone. Numerical
 simulations generally support the results of the analytical study.
\end{abstract}


\section{Introduction}

For many years, it was thought that radiative losses via X-ray emission
in clusters of galaxies leads to a substantial gas inflow, which was
called a ``cooling flow'' \citep[][and references therein]{fab94}.
However, X-ray spectra taken with {\it ASCA} and {\it XMM-Newton} fail
to show line emission from ions having intermediate or low temperatures,
implying that the cooling rate is at least 5 or 10 times less than that
previously assumed
\citep[e.g.][]{ike97,mak01,pet01,tam01,kaa01,mat02}. {\it Chandra}
observations have also confirmed the small cooling rates
\citep*[e.g.][]{mcn00,joh02,ett02,bla03}.

These observations suggest that a gas inflow is prevented by some heat
sources that balance the radiative losses. There are two popular ideas
about the heating sources. One is energy injection from a central AGN of
a cluster \citep*{tuc83,boh88,rep87,bin95,sok01,cio01,boh02,
chu02,sok02,rey02,kai03}.  Recent {\it Chandra} observations show that
AGNs at cluster centers actually disturb the intracluster medium (ICM)
around them \citep*{fab00,mcn00,bla01,mcn01,maz02,fuj02,joh02,kem02},
although some of them were already discovered by {\it ROSAT}
\citep{boh93,hua98}. Numerical simulations suggest that buoyant bubbles
created by the AGNs mix and heat the ambient ICM to some extent
\citep*{chu01,qui01,sax01,bru02,bas03}. The other possible heat source
is thermal conduction from the hot outer layers of clusters
\citep{tak79,tak81,tuc83,fri86,gae89,boh89a,spa92,sai99,nar01}.

However, it has already been known that the ICM heating by AGNs or
thermal conduction has problems.  For the AGN heating, the efficiency of
the heating must be quite high \citep*{fab02}. Moreover, the
intermittent activity of an AGN makes the temperature profile of the
host cluster irregular, which is inconsistent with observations
\citep{bri03}. For the thermal conduction, stability is the most serious
problem; either the observed temperature gradient disappears or the
conduction has a negligible effect relative to radiative cooling
\citep{bre88,bri03,sok03}. Moreover, thermal conduction alone cannot
sufficiently heat the central regions of some clusters
\citep{voi02,zak03}. Although a ``double heating model'' that
incorporates the effects of simultaneous heating by both the central AGN
and thermal conduction may alleviate the stability problem
\citep{rus02}, \citet{bri03} indicates that the conductivity must still
be about $0.35\pm 0.10$ times the Spitzer value.

In this paper, we consider another natural heating source. In the ICM,
fluid turbulence is generated by substructure motion or cluster
mergers. From numerical simulations, \citet*{nag03} showed that the
turbulent velocities in the ICM is about 20\%--30\% of the sound speed
even when a cluster is relatively relaxed. Such turbulence generates
acoustic waves in the ICM. Compressive characters of the acoustic waves
with a relatively large amplitude inevitably lead to the the steepening
of the wave fronts to form shocks. As a result, the waves can heat the
surrounding ICM through the shock dissipation. A similar heating
mechanism has also been proposed in the solar corona; the waves are
excited by granule motions of surface convection
\citep*{ost61,ulm71,mcw75}. The idea of wave heating in the ICM was
proposed by \citet{pri89}, but the study was limited to
order-of-magnitude estimates. In this paper, we study the wave heating
by an analytical model and numerical simulations. We use cosmological
parameters of $\Omega_0=0.3$, $\lambda=0.7$, and $H_0=70\:\rm km\:
s^{-1}\: Mpc^{-1}$ unless otherwise mentioned.

\section{Analytical Approach}
\subsection{Models}

In the ICM the magnetic pressure is generally negligible against the gas
pressure \citep{sar86}. Therefore, acoustic waves (strictly speaking,
fast mode waves in high-$\beta$ plasma) could carry much larger amount
of energy than other modes of magnetohydrodynamical waves.  We expect
that turbulence in the ICM excites acoustic waves that propagate in
various directions. In this paper, we focus on the acoustic waves
traveling inward, which play an important role in the heating of the
cluster center.  These waves, having a relatively large but finite
amplitude, eventually form shocks to shape sawtooth waves ({\sf
N}-waves) and directly heat the surrounding ICM by dissipation of their
wave energy. We adopt the heating model for the solar corona based on
the weak shock theory \citep{suz02,ss72}. In this section, we assume
that a cluster is spherically symmetric and stationary. The equation of
continuity is
\begin{equation}
\label{eq:cont}
 \dot{M}=-4\pi r^2\rho v\:,
\end{equation}
where $\dot{M}$ is the mass accretion rate, $r$ is the distance from the
cluster center, $\rho$ is the gas density, and $v$ is the gas velocity.
The equation of momentum conservation is
\begin{equation}
\label{eq:motion}
 v\frac{dv}{dr} = -\frac{GM(r)}{r^2}-\frac{1}{\rho}\frac{dp}{dr}
-\frac{1}{\rho c_s \{1+[(\gamma+1)/2]\alpha_w\}}
\frac{1}{r^2}\frac{d}{dr}(r^2 F_w)
\end{equation}
where $G$ is the gravitational constant, $M(r)$ is the mass within
radius $r$, $p$ is the gas pressure, $c_s$ is the sound velocity,
$\gamma (=5/3)$ is the adiabatic constant, and $\alpha_w$ is the wave
velocity amplitude normalized by the ambient sound velocity
($\alpha_w=\delta v_w/c_s$). For the actual calculations, we ignore the
term $vdv/dr$ because the velocity is much smaller than the sound
velocity except for the very central region of a cluster where the weak
shock approximation is not valid ($\alpha_w\gtrsim 1$; see
\S\ref{sec:results}). The wave energy flux, $F_w$, is given by
\begin{equation}
\label{eq:Fw}
 F_w = -\frac{1}{3}\rho c_s^3 
\alpha_w^2 \left(1+\frac{\gamma+1}{2}\alpha_w\right) \:.
\end{equation}
Note that the sign of equation~(\ref{eq:Fw}) is the opposite of
equation~(7) of \citet{suz02}, because we consider waves propagating
inwards contrary to those in \citet{suz02}. The energy equation is
written as
\begin{equation}
\label{eq:energy}
\rho v \frac{d}{dr}\left(\frac{1}{2}v^2
+\frac{\gamma}{\gamma-1}\frac{k_B T}{\mu m_H}\right)
+\rho v \frac{G M(r)}{r^2}
+\frac{1}{r^2}\frac{d}{d r}[r^2(F_w+F_c)]
+n_e^2\Lambda(T,Z)=0 \:,
\end{equation}
where $k_B$ is the Boltzmann constant, $T$ is the gas temperature, $\mu
(=0.61)$ is the mean molecular weight, $m_H$ is the hydrogen mass, $n_e$
is the electron number density, and $\Lambda$ is the cooling
function. The term $\nabla\cdot\mbox{\boldmath $F$}_w$ indicates the
heating by the dissipation of the waves. We adopt the classical form of
the conductive flux for ionized gas,
\begin{equation}
\label{eq:Fc}
 F_c = -f_c \kappa_0 T^{5/2}\frac{dT}{dr}
\end{equation}
with $\kappa_0 = 5\times 10^{-7}$ in cgs units. The factor $f_c$ is the
ratio of actual thermal conductivity to the classical Spitzer
conductivity. The cooling function is a function of temperature $T$ and
metal abundance $Z$, and is given by
\begin{eqnarray}
 \Lambda(T,Z) &=& 2.1\times 10^{-27}
\left(1+0.1\frac{Z}{Z_\sun}\right)\left(\frac{T}{\rm K}\right)^{-0.5}
\\
 & &+\;  8.0\times 10^{-17}
\left(0.04+\frac{Z}{Z_\sun}\right)\left(\frac{T}{\rm K}\right)^{-1.0}
\end{eqnarray}
in units of $\rm ergs\: cm^3\: s^{-1}$. This is an empirical formula
derived by fitting to the cooling curves calculated by \citet{boh89}.
We assume that wave injection takes place at radii far distant from the
cluster center, and thus there is no source term of waves in
equation~(\ref{eq:energy}).

The equation for the evolution of shock wave amplitude is given by
\begin{equation}
\label{eq:wave}
 \frac{d\alpha_w}{dr}=\frac{\alpha_w}{2}\left[
-\frac{1}{p}\frac{dp}{dr}
+\frac{2(\gamma+1)\alpha_w}{c_s \tau}-\frac{2}{r}
-\frac{1}{c_s}\frac{d c_s}{dr}
\right]\;,
\end{equation}
where $\tau$ is the period of waves, which we assume to be constant
\citep{suz02}. We give the period by $\tau = \lambda_0/c_{s0}$, where
$c_{s0}$ is the sound velocity at the average temperature of a cluster
($T_{\rm av}$), and $\lambda_0$ is the wave length given as a
parameter. The second term of the right side of equation~(\ref{eq:wave})
denotes dissipation at each shock front of the {\sf N}-waves. We note
that the sign of the term is the opposite of equation~(6) of
\citet{suz02}, because we consider waves propagating inwards contrary to
those in \citet{suz02}.

For the mass distribution of a cluster, we adopt the NFW profile
\citep*{nav97}. The mass profile is written as
\begin{equation}
\label{eq:NFW}
 M(r) \propto \left[\ln \left(1+\frac{r}{r_s}\right)
-\frac{r}{r_s (1+r/r_s)}
\right]\:,
\end{equation}
where $r_s$ is the characteristic radius of the cluster.  The
normalization can be given by $M(r_{\rm vir})=M_{\rm vir}$, where
$r_{\rm vir}$ and $M_{\rm vir}$ are the virial radius and mass of a
cluster, respectively. We ignore the self-gravity of the ICM.

\subsection{Results}
\label{sec:results}

We show that our model can reproduce observed ICM density and
temperature profiles of clusters. We choose A1795 and Ser~159--03
clusters to be compared with our model predictions. \citet{zak03} showed
that thermal conduction alone can explain the density and temperature
profiles for A1795; $f_c=0.2$ is enough and other heat sources are not
required. On the other hand, the profiles for Ser~$159-03$ cannot be
reproduced by thermal conduction alone \citep{zak03}. The parameters of
the mass profiles for the clusters are the same as those adopted by
\citet{zak03} and are shown in Table~\ref{tab:mass}. The concentration
parameter of a cluster, $c=r_{\rm vir}/r_s$ is given by
\begin{equation}
 \label{eq:r_vir}
c=\frac{1}{r_s}\left[\frac{3\: M_{\rm vir}}
{4\pi\: 200\:\rho_{\rm crit}}\right]^{1/3}\:,
\end{equation}
where $\rho_{\rm crit}$ is the critical density of the universe. We fix
the metal abundance profiles. For A1795, we assume $Z(r)=0.8
\exp(-r/170{\rm\: kpc})\: Z_\sun$ \citep{ett02}, and for Ser~159--03, we
assume $Z(r)=0.51 \exp(-r/171{\rm\: kpc})\: Z_\sun$ \citep{kaa01}.

We carry out the modeling of ICM heating as follows. First, we select
values of $f_c$, $\dot{M}$, and $\lambda_0$. Then, we set the boundary
conditions of the equations~(\ref{eq:cont}), (\ref{eq:motion}),
(\ref{eq:energy}), and (\ref{eq:wave}) at $r_i=1$~kpc, that is, well
inside the central cD galaxy. From $r=r_i$, we integrate the equations
outward and compare the model profiles of $n_e(r_i)$ and $T(r_i)$ with
the data. While we fix the value of $\alpha_w(r_i)$, we adjust
$n_e(r_i)$ and $T(r_i)$ to be consistent with the observed profiles. We
restrict ourselves to a comparison by eye, since neither the data nor
the models are reliable enough for a detailed $\chi^2$ fit. If we do not
have satisfactory fits, we change the values of $f_c$, $\dot{M}$, and
$\lambda_0$ and repeat the process. We show the values of $f_c$,
$\dot{M}$, and $\lambda_0$ that we finally adopted in
Table~\ref{tab:fit}, and briefly summarize how the results depend on the
choice of them as follows.

For A1795, we choose $f_c = 2\times 10^{-3}$, because \citet{zak03} have
already shown that ICM heating only by thermal conduction with $f_c\sim
0.2$ is consistent with the observations. In this study, we will show
that even when $f_c$ is much smaller than 0.2, the observed profiles can
be reproduced if wave heating is included. However, we found that if
$f_c$ is too small, the obtained temperature profile is too steep to be
consistent with the observation. For Ser~159--03, we adopt $f_c = 0.2$,
which is suggested by \citet{nar01} in a turbulent MHD medium. If we
take $f_c$ much smaller than this, the model cannot reproduce the
relatively flat temperature distribution observed in this cluster.

For mass accretion rates $\dot{M}$, we take about 1/10 times the value
claimed before the {\it Chandra} and {\it XMM-Newton} era. For A1795,
$\dot{M}\sim 500\: M_\sun\rm\: yr^{-1}$ ($h=0.5$) was reported
\citep*{edg92,per98}. Thus, we adopt $\dot{M}=50\: M_\sun\rm\: yr^{-1}$,
which is consistent with a recent {\it XMM-Newton} observation
\citep[$<150\: M_\sun\rm\: yr^{-1}$;][]{tam01}. For Ser~159--03,
$\dot{M}\sim 300\: M_\sun\rm\: yr^{-1}$ ($h=0.5$) was reported
\citep*{whi97,all97}. Thus, we adopt $\dot{M}=30\: M_\sun\rm\: yr^{-1}$.
We note that if we assume that wave heating is effective and that
$\dot{M}$ is much smaller than the above values, we cannot reproduce
both density and temperature profiles obtained by X-ray observations; we
get too high temperature and too low density.

Typical wave length, $\lambda_0$, should be comparable to the typical
eddy size of turbulence in ICM.  From numerical simulations,
\citet*{roe99} showed that the typical eddy size is the core scale of a
cluster. Thus, we take $\lambda_0=100$~kpc for A1795. For Ser~159--03,
we use a smaller value of $\lambda_0=70$~kpc because of its small mass
(Table~\ref{tab:mass}). Smaller $\lambda_0$ means a smaller distance
that waves propagate before dissipation.

Among three of the parameters for the boundary conditions at $r=r_i$
($\alpha_w$, $n_e$, and $T$), we fix $\alpha_w=3$ to reduce the number
of fitting parameters. If we assume much smaller $\alpha_w$, wave
heating becomes negligible. On the other hand, if we assume much larger
$\alpha_w$, the region where the weak shock approximation is invalid
($\alpha_w\gtrsim 1$) extends.

\begin{figure}\epsscale{0.45}
\plotone{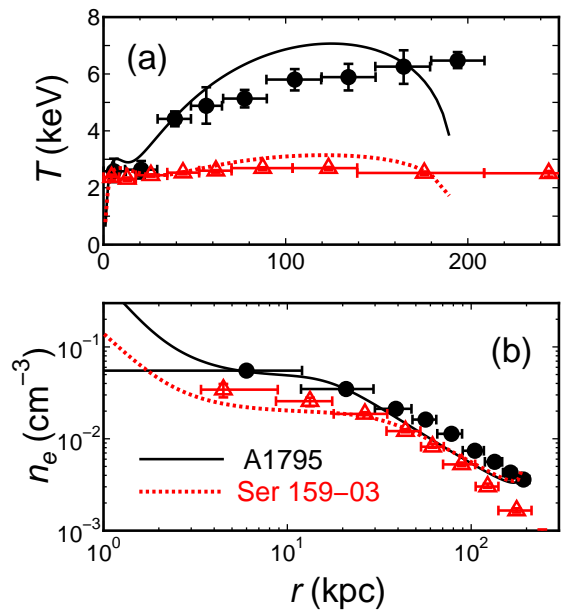}\caption{(a) Modeled temperature and (b)
 density profiles for A1795 (solid lines) and Ser~159--03 (dotted
 lines). Filled dots and empty triangles are the {\it Chandra} data for
 A1795 and Ser~159--03, respectively. \label{fig:Tn}}
\end{figure}

Figure~\ref{fig:Tn} shows the model fits for the two clusters. The
boundary conditions are presented in Table~\ref{tab:fit}. The
temperature $T(r_i)$ is especially required to be fine-tuned for the
fit. We use the {\it Chandra} data of A1795 obtained by \citet{ett02}
and the {\it XMM-Newton} data of Ser~159--03 obtained by
\citet{kaa01}. The {\it XMM-Newton} data of A1795 are also obtained by
\citet{tam01} and they are similar to those obtained by \citet{ett02},
although the former is not deprojected contrary to the latter. The good
agreement between the model and the data suggests that wave heating is a
promising candidate of the mechanism that solves the cooling flow
problem. In Figure~\ref{fig:Tn}, densities go to infinity and
temperatures go down to zero at $r\sim 200$~kpc. This suggests that
waves injected outside of this radius cannot reach the cluster center.

\begin{figure}\epsscale{0.45}
\plotone{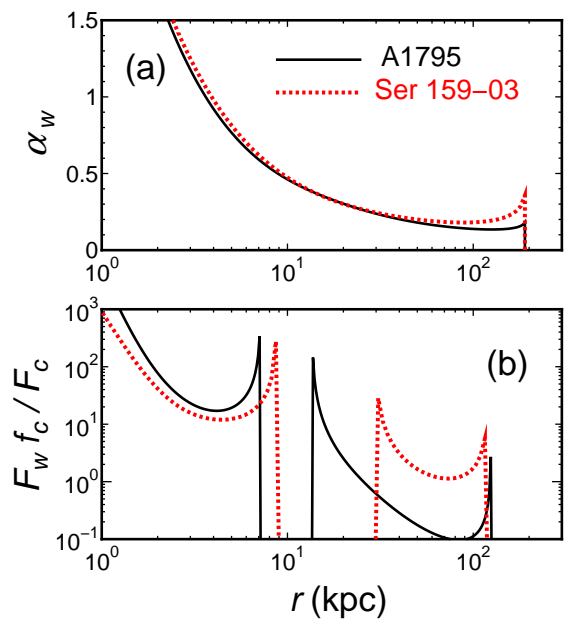} \caption{(a) Wave amplitudes and (b) the ratio of heat
flux by waves to that by thermal conduction for A1795 (solid lines) and
Ser~159--03 (dotted lines). \label{fig:aF}}
\end{figure}

In Figure~\ref{fig:aF}a, we present the wave velocity amplitude
normalized by the sound velocity ($\alpha_w$). As the {\sf N}-waves
propagate into the central regions of the cluster, $\alpha_w$ increases
rapidly. This is mainly because of the geometrical convergence to the
cluster center, whereas the total wave luminosity (= energy flux times
$r^2$) mostly dissipates through the inward propagation in itself.
Since $\alpha_w>1$ at $r\lesssim 4$~kpc for both A1795 and Ser~159--03,
the results may not be quantitatively correct there.  We note that the
gas velocity for the region of $r\gtrsim 4$~kpc is very small compared
with the sound velocity and thus our ignorance of $vdv/dr$ in
equation~(\ref{eq:motion}) is justified.

In Figure~\ref{fig:aF}b, the ratio $F_w f_c/F_c$ is presented. The gaps
at $\sim 10$~kpc reflect $F_c<0$. Assume that an observer made
observations of the model clusters and the temperature distributions
were exactly measured. If the observer {\it assumed} the classical
conductivity, the heat flux measured by the observer should be $F_c/f_c$
($f_c<1$) because of the definition of $F_c$
(equation~[\ref{eq:Fc}]). Figure~\ref{fig:aF}b shows that the observer
would measure an X-ray emission much larger than that predicted by the
classical thermal conduction ($F_w f_c/F_c>1$) if the energy swallowed
by the black hole at the cluster center is small. Such large X-ray
emissions have actually been estimated in some clusters
\citep{voi02}. Wave heating model can account for the observations
without the help of heating by AGNs.

\section{One-Dimensional Numerical Simulations}

\subsection{Models}

In order to be compared with the results in the previous section, we
performed one-dimensional numerical simulations. We solve the following
equations:
\begin{equation}
 \frac{\partial \rho}{\partial t}
+ \frac{1}{r^2}\frac{\partial}{\partial r}(r^2 \rho v)=0 \;,
\end{equation}
\begin{equation}
 \frac{\partial (\rho v)}{\partial t}
+\frac{1}{r^2}\frac{\partial}{\partial r}(r^2 \rho v^2)
=-\rho \frac{G M(r)}{r^2}-\frac{\partial p}{\partial r}
\;,
\end{equation}
\begin{equation}
\label{eq:energy2}
 \frac{\partial e}{\partial t}+\frac{1}{r^2}[r^2 v(p+e)]
=\frac{1}{r^2}\frac{\partial}{\partial r}
\left[r^2 \kappa(T)\frac{\partial T}{\partial r}\right]
-n_e^2 \Lambda(T)-\rho u \frac{G M(r)}{r^2} \;,
\end{equation}
where the total energy is defined as $e=p/(\gamma-1)+\rho v^2/2$, and
$\kappa(T)=\kappa_0 T^{5/2}$. We ignore the self-gravity of ICM. For
numerical simulations, we adopt the cooling function based on the
detailed calculations by \citet{sut93},
\begin{equation}
\label{eq:cool}
 n_e^2 \Lambda = [C_1 (k_B)^\alpha + C_2 (k_B T)^\beta + C_3]n_i n_e\:,
\end{equation}
where $n_i$ is the ion number density and the units for $k_B T$ are
keV. For an average metallicity $Z=0.3\: Z_\sun$ the constants in
equation (\ref{eq:cool}) are $\alpha=-1.7$, $\beta=0.5$, $C_1=8.6\times
10^{-3}$, $C_2=5.8\times 10^{-2}$, and $C_3=6.4\times 10^{-2}$, and we
can approximate $n_i n_e=0.704(\rho/m_H)^2$. The units of $\Lambda$ are
$10^{-22}\:\rm ergs\: cm^3$ \citep{rus02}. Note that
equation~(\ref{eq:energy2}) does not include the energy dissipation term
contrary to equation~(\ref{eq:energy}). This is because the energy
dissipation at shocks is included automatically in numerical simulations
if shocks are resolved.

The hydrodynamic part of the equations is solved by a second-order
advection upstream splitting method (AUSM) based on \citet[][see also
\citealt{wad01}]{lio93}. We use 500 unequally spaced meshes in the
radial coordinate to cover a region with a radius of 300~kpc. The inner
boundary is set at $r=1$~kpc. The innermost mesh has a width of $\sim
15$~pc, and the width of the outermost mesh is $\sim 3$~kpc. The
following boundary conditions are adopted:
\begin{enumerate}
 
\item Variables except velocity have zero gradients at the center.

\item The inner edge is assumed to be a perfectly reflecting
point.

\item The density and pressure at the outermost mesh are equal to
specified values.
 
\end{enumerate}
Waves are injected at the outermost mesh as
\begin{equation}
 v(t)=\alpha_{w0}c_{s0}
\sin\left(\frac{2\pi c_{s0} t}{\lambda_0}\right)\:,
\end{equation}
where $\alpha_{w0}$ is the parameter.

\subsection{Numerical Results}

The mass distribution we assumed is the same as that of A1795 in
Table~\ref{tab:mass} except for $T_{\rm av}$. We assume that the ICM is
isothermal and in pressure equilibrium at $t=0$. For the NFW profile
(equation~[\ref{eq:NFW}]), the gas initial density profile is written as
\begin{equation}
 \rho(r) = \rho_0 \exp[-B f(r/r_s)] \;,
\end{equation}
where
\begin{equation}
 B = \frac{1}{m(1)}\frac{G\mu m_H M(r_s)}{r_s k_B T_{\rm av}} \;,
\end{equation}
\begin{equation}
 m(x) = \ln(1+x)-\frac{x}{1+x} \;,
\end{equation}
\begin{equation}
 f(x) = 1-\frac{1}{x}\ln(1+x) \;
\end{equation}
\citep*{sut98}. The initial ICM temperature is $T_{\rm av}$. The density
and pressure of the outer boundary are fixed at the initial values. We
finish the calculations when the temperature of some of the meshes
becomes zero because we do not treat mass dropout from the hot ICM. We
define the time as $t_{\rm cool}$. If the temperature does not go to
zero, we finish the calculations at $t=7$~Gyr. In this section, we set
$n_e(0)=0.017\rm\: cm^{-3}$, $T_{\rm av}=7$~keV, and
$\lambda_0=100$~kpc. Other model parameters are shown in
Table~\ref{tab:sim}.

\begin{figure}\epsscale{0.5}
\plotone{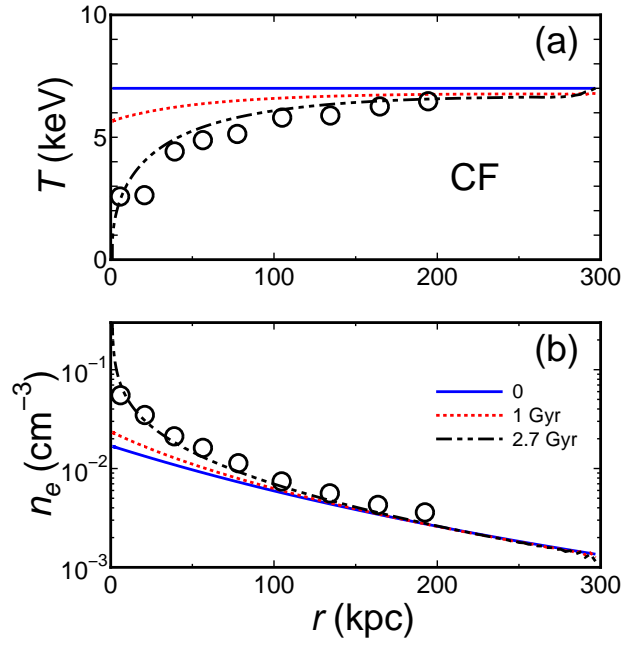} \caption{(a) Temperature and (b) density
 profiles for Model~CF. Open circles are the {\it Chandra} data for A1795
 \citep{ett02}. \label{fig:CF}}
\end{figure}

\begin{figure}\epsscale{0.5}
\plotone{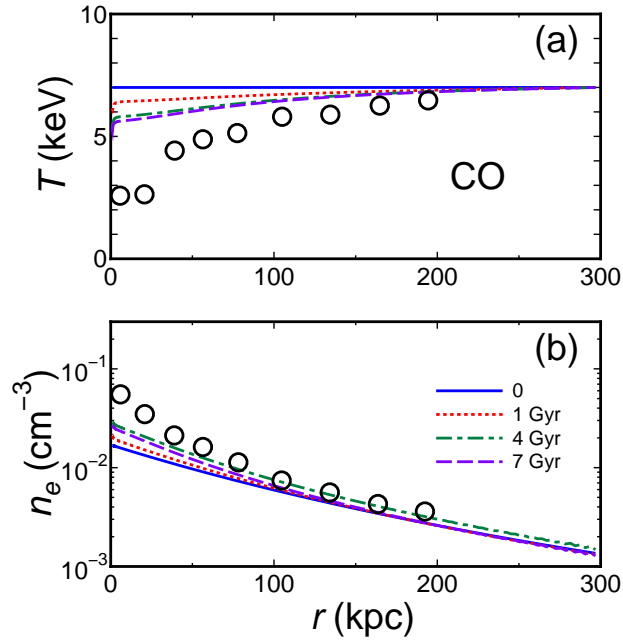} \caption{The same as Figure~\ref{fig:CF} but for
Model~CO\label{fig:CO}}
\end{figure}

Figures~\ref{fig:CF} and~\ref{fig:CO} show the temperature and density
profiles for Models~CF and~CO, respectively. For reference, the data of
A1795 are shown \citep{ett02}, although the detailed comparison is
premature.  Contrary to Figure~\ref{fig:Tn}, we use a linear scale for
the distance from the cluster center for both temperature and density
profiles to see shock structures. Model~CF is a genuine cooling flow
model; the temperature goes to zero at the cluster center at $t_{\rm
cool}=2.7$~Gyr. In Model~CO, conduction dominates cooling and the
solution is stable for a long time. The suppression factor of the
conductivity, $f_c=0.2$, in Model~CO is the same as that of
\citet{zak03}. The temperature gradient at the cluster center is less
steep than that obtained by \citet{zak03}. This may be because of the
differences of the boundary conditions or the initial conditions.

\begin{figure}\epsscale{0.5}
\plotone{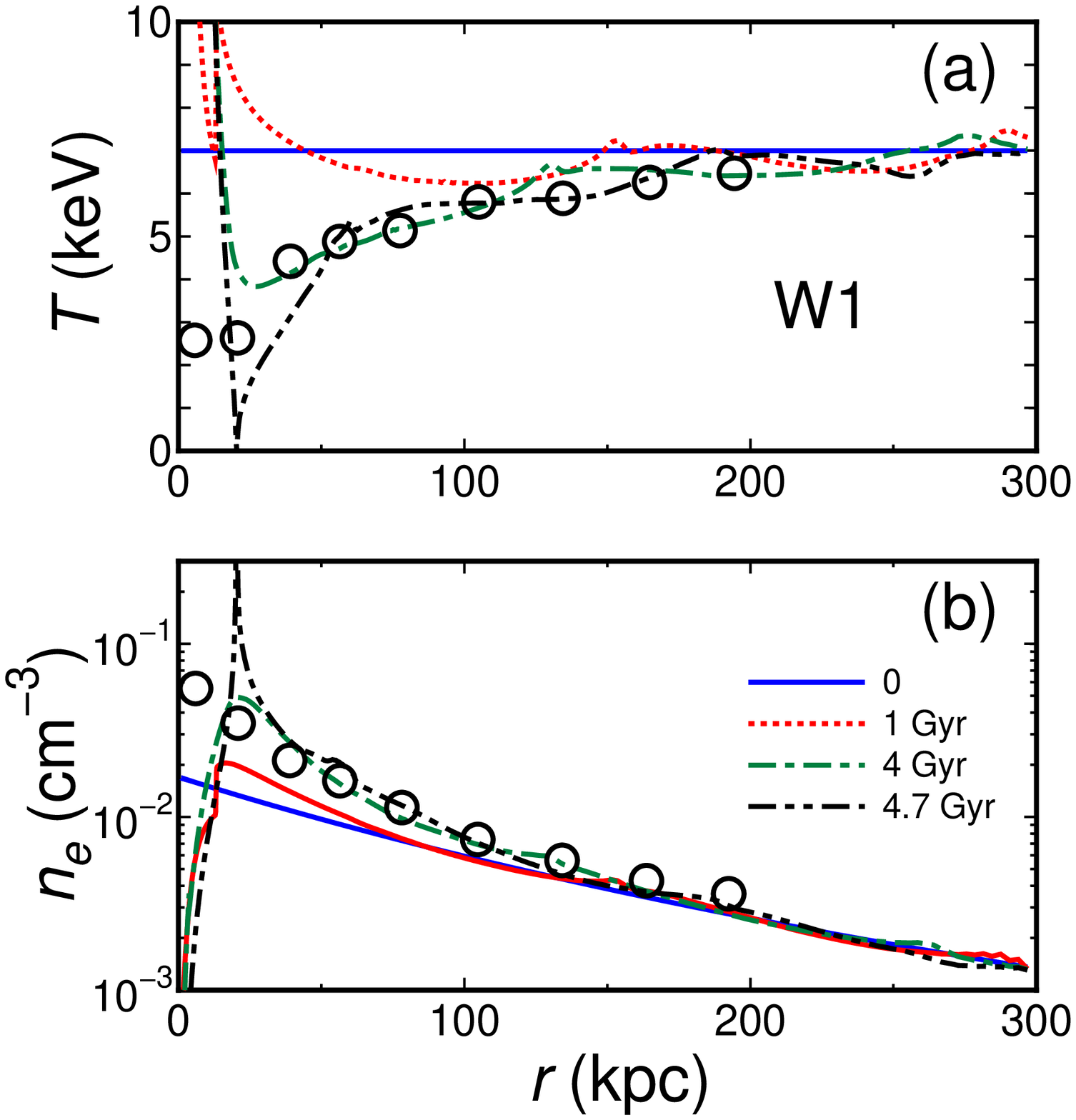} \caption{The same as Figure~\ref{fig:CF} but for
Model~W1\label{fig:W1}}
\end{figure}

\begin{figure}\epsscale{0.5}
\plotone{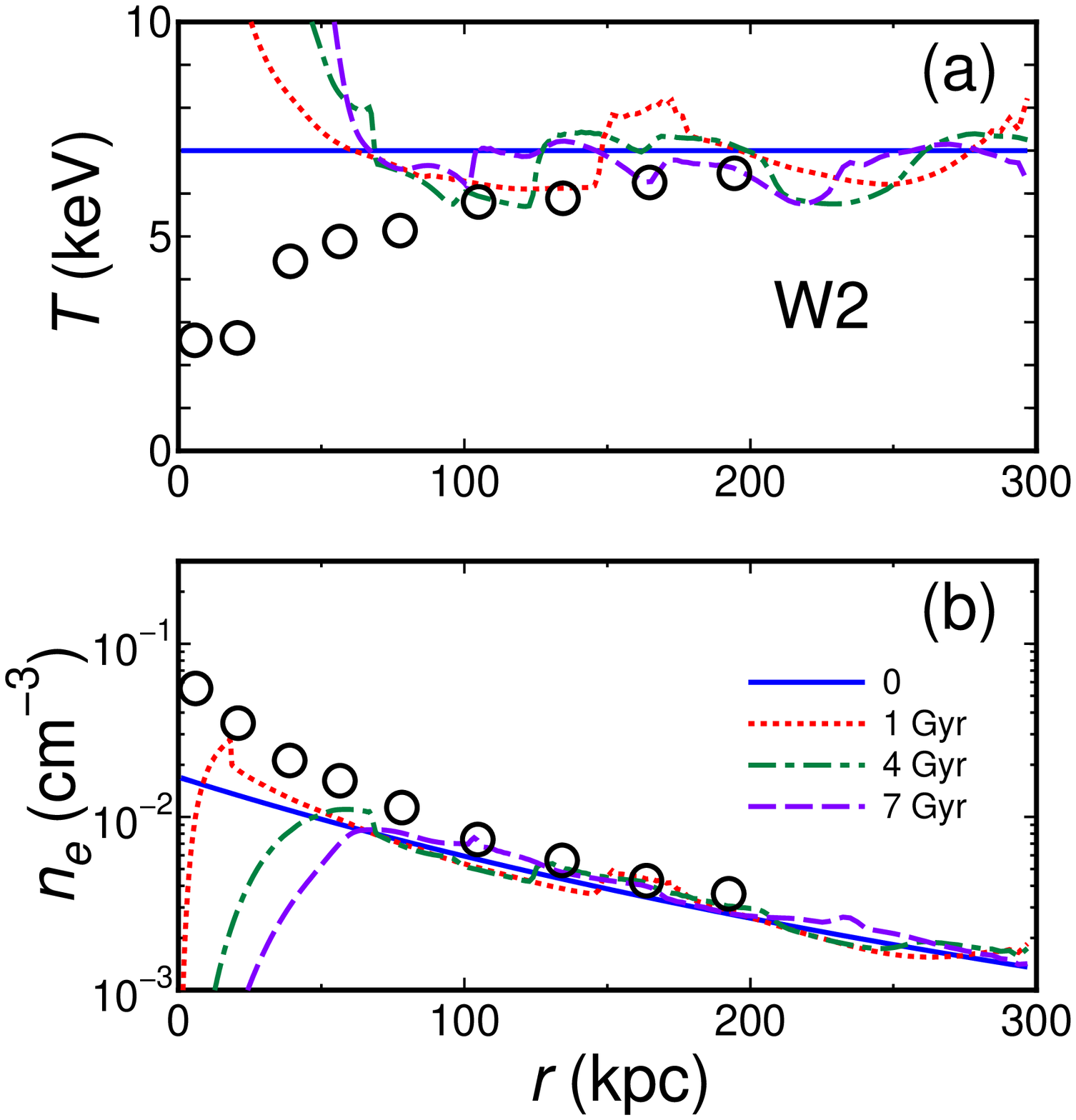} \caption{The same as Figure~\ref{fig:CF} but for
Model~W2\label{fig:W2}}
\end{figure}

Figures~\ref{fig:W1} and~\ref{fig:W2} show the temperature and density
profiles for Models~W1 and~W2, respectively. As we predicted in
Figure~\ref{fig:aF}, waves are amplified in the central region. A shock
is seen at $r=13$~kpc in Model~W1. Both temperature and density diverge
at the cluster center because these are one-dimensional simulations and
waves focus on the cluster center. In the outer region, the shape of
waves is complicated. This is because we adopted the reflection
condition at the inner boundary and the inward and outward waves
interact. For Model~W1, $t_{\rm cool}=4.7$~Gyr, and for Model~W2,
$t_{\rm cool}>7$~Gyr. They are much larger than $t_{\rm cool}$ for
Model~CF. This means the wave heating is effective.

In Models~M1 and~M2, we include both thermal conduction and waves. The
results of Model~M1 and Model~M2 are almost the same as those of
Model~W1 and~CO, respectively. The cooling time-scale $t_{\rm cool}$ of
Model~M1 is shorter than that of Model~W1, because shocks are weakened
by the thermal conduction. In model~M2, the conduction dominates the
wave heating.

\section{Discussion}

Our analytical model and numerical simulations show that acoustic waves
are amplified at the cluster center and can heat the cluster core. One
should note that our assumption, that is, the spherical symmetry, could
affect the amplification quantitatively. For more realistic modeling, we
should consider that real clusters are not exactly spherically
symmetric. However, \citet{pri89} indicated that even if a cluster is
not spherically symmetric, the lower temperature and smaller sound
velocity at the cluster center should have waves focus on the center. In
order to study this effect, we need to perform high-resolution
multi-dimensional numerical simulations (Wada, Fujita, \&, Suzuki 2003,
in preparation). Since the focusing effect depends on the temperature
gradient, it may solve the fine-tuning problem of the heating of cluster
cores. If the cooling dominates heating, the temperature at the cluster
center decreases and the temperature gradient in the core
increases. This strengthens the focusing effect and wave heating becomes
more efficient. Multi-dimensional simulations have another benefit; it
is free from the inner boundary conditions that we set in our
one-dimensional simulations.

If waves are actually responsible for the heating in cluster cores, weak
shocks should be observed there in some clusters. In our simulations,
waves are amplified at $r\lesssim 10-50$~kpc (Figures~\ref{fig:W1}
and~\ref{fig:W2}). Thus, if the wave length is large ($\gtrsim
100$~kpc), the shocks are not necessarily observed in all
clusters. However, as pointed out by \citet{chu03}, the passing of waves
may cause gas-sloshing around cluster cores. Observations of fine
structures in many cluster cores will be useful to understand the
heating mechanism there.

In the central regions of clusters, turbulence generated by interaction
of amplified waves from various directions could be observed as
broadened metal lines by future X-ray observatories. Observational
studies of optical emission lines revealed that there is warm gas at
some cluster centers and that the velocity widths range from 100 to
$1000\rm\: km\: s^{-1}$ \citep*{hu85,joh87,hec89}. These warm gas may be
embodied in and move with the hot turbulent ICM \citep[see][]{loe90}.

\section{Conclusions}

Through analytical and numerical approaches, we have shown that acoustic
waves generated by turbulence in the ICM in the outer region of a
cluster can effectively heat the central part of the cluster. The heat
flux by the waves may exceed that by thermal conduction. The process
presented here is phenomenologically analogous to the collapse of a
"tsunami" (a seismic sea wave) at seashore owing to the change of depth
of the sea. As in tsunamis, even if the waves have small amplitude at
their origin, they could bring huge damage at a distant point, namely
the cluster center.  Of course, one should note that tsunamis are
gravity-driven waves and not acousitc waves. In the analytical studies,
we have obtained time-independent solutions and compared the predicted
density and temperature profiles with the observed ones; they are
consistent with each other. In general, we have confirmed the results
obtained by the analytical studies by one-dimensional numerical
simulations. Since we assumed that a cluster is spherically symmetric
and the assumption leads to artificial focusing of waves, one should
take the quantitative results with care. However, it has been indicated
that even if a cluster is not spherically symmetric, waves are focused
by the temperature gradient at the cluster center. Thus, it is
worthwhile to study the wave heating by multi-dimensional analyses.

\acknowledgments

Y.~F.\ and K.~W were supported in part by a Grant-in-Aid from the
Ministry of Education, Culture, Sports, Science, and Technology of Japan
(Y.~F.:14740175; K.~W.:15684003). T.~K.~S. is financially supported by
the JSPS Research Fellowship for Young Scientists, grant 4607.

\newpage

\begin{deluxetable}{ccccc}
\tablecaption{Cluster Parameters  \label{tab:mass}}
\tablewidth{0pt}
\tablehead{
\colhead{Cluster} & $M_{\rm vir}$ & $T_{\rm av}$ & 
$r_s$  & $c$  
\\
  & ($10^{14}\: M_{\sun}$) & (keV) & 
(Mpc)  & 
}
\startdata
A1795        & 12 & 7.5 & 0.46 & 4.2 \\
Ser~159--03 & 2.6 & 2.7 & 0.31 & 4.7 \\
\enddata

\end{deluxetable}

\begin{deluxetable}{ccccccc}
\tablecaption{Model Parameters  \label{tab:fit}}
\tablewidth{0pt}
\tablehead{
\colhead{Cluster} & $f_c$ & $\dot{M}$ & $\lambda_0$ & 
$\alpha_w(r_i)$  & $n_e(r_i)$ & $T(r_i)$   
\\
  &  & ($M_\sun \rm\: yr^{-1}$) &  (kpc) &
  & ($\rm cm^{-3}$) & (keV)
}
\startdata
A1795       & $2\times 10^{-3}$ & 50 & 100 & 3 & 0.5  & 0.6213 \\
Ser~159--03 & 0.2               & 30 &  70 & 3 & 0.14 & 0.780 \\
\enddata

\end{deluxetable}

\begin{deluxetable}{cccc}
\tablecaption{Model Parameters  \label{tab:sim}}
\tablewidth{0pt}
\tablehead{
\colhead{Model} & $\alpha_{w0}$ & $f_c$ & $t_{\rm cool}$ (Gyr)
}
\startdata
CF       & 0   & 0     & 2.7      \\
CO       & 0   & 0.2   & \nodata  \\
W1       & 0.1 & 0     & 4.7      \\
W2       & 0.2 & 0     & \nodata  \\
M1       & 0.1 & 0.002 & 4.3      \\
M2       & 0.1 & 0.2   & \nodata  \\
\enddata
\tablecomments{No data for $t_{\rm cool}$ mean $t_{\rm cool}>7$~Gyr.}
\end{deluxetable}

\newpage

\newpage

\newpage

\end{document}